\documentclass[10pt,journal,compsoc]{IEEEtran}

\ifCLASSOPTIONcompsoc
\usepackage[nocompress]{cite}
\else
\usepackage{cite}
\fi

\ifCLASSINFOpdf

\else

\fi

\usepackage{textcomp}
\usepackage{verbatim}
\usepackage{amsfonts}
\usepackage{amsmath}
\usepackage{array}
\usepackage{balance}
\usepackage{multicol}
\usepackage{multirow}
\usepackage{color}
\usepackage{epsfig}
\usepackage[numbers]{natbib}
\usepackage{xspace}
\usepackage{booktabs}
\usepackage{bm}
\usepackage{enumitem}
\usepackage{diagbox}
\usepackage{url}
\usepackage{changepage}
\usepackage{textcomp}
\usepackage{graphicx}
\graphicspath{ {./fig/} }
\usepackage{color}
\usepackage{bm}
\usepackage{amssymb}
\usepackage{algorithm}
\usepackage{algorithmic}
\usepackage{threeparttable}
\usepackage{multirow}
\usepackage{subfigure}
\usepackage{mathrsfs}

\DeclareMathAlphabet{\mathcal}{OMS}{cmsy}{m}{n}

\newcommand{\hide}[1]{} 
\newcommand{\vpara}[1]{\noindent\textbf{#1 }}

\begin{document}

\title{Who’s Next: Rising Star Prediction via Diffusion of User Interest in Social Networks}

\def\model#1{RiseNet}
\def\gnnm#1{Graph Neural Network Module}
\def\rnnm#1{Recurrent Neural Network Module}

\author{Xuan Yang,
	Yang Yang,
	Jintao Su,
	Yifei Sun,
	Shen Fan,
	Zhongyao Wang,
	Jun  Zhang,
	Jingmin  Chen
	\IEEEcompsocitemizethanks{\IEEEcompsocthanksitem 
		Xuan Yang, Yang Yang and Jintao Su are with the College of Computer Science and Technology, 
		Zhejiang University, China. E-mail: {xuany, yangya, jtsu}@zju.edu.cn.
		Yang Yang is the corresponding author. 
        \IEEEcompsocthanksitem Yifei Sun is with the College of Aeronautics and Astronautics, Zhejiang University, China. E-mail: sunyf@zju.edu.cn.
 		\IEEEcompsocthanksitem Zhongyao Wang, Shen Fan, Jun Zhang, Jingmin Chen are with the Alibaba Group, China. E-mail: {zhongyao.wangzy, fanshen.fs}@alibaba-inc.com, socialnetwork@gmail.com, jingmin.cjm@alibaba-inc.com.
		}
	\thanks{Manuscript received ***; revised ***.}
}

\markboth{Journal of IEEE Transactions on Knowledge and Data Engineering,~Vol.~*, No.~*, **}%
{Shell \MakeLowercase{\textit{et al.}}: Bare Demo of IEEEtran.cls for Computer Society Journals}

\IEEEtitleabstractindextext{%

\begin{abstract}
Finding items with potential to increase sales is of great importance in online market. In this paper, we propose to study this novel and practical problem: rising star prediction. We call these potential items \emph{Rising Star}, which implies their ability to rise from low-turnover items to best-sellers in the future. Rising stars can be used to help with unfair recommendation in e-commerce platform, balance supply and demand to benefit the retailers and allocate marketing resources rationally. 
Although the study of rising star can bring great benefits, it also poses challenges to us. The sales trend of rising star fluctuates sharply in the short-term and exhibits more contingency caused by some external events (e.g., COVID-19 caused increasing purchase of the face mask) than other items, which cannot be solved by existing sales prediction methods.

To address above challenges, in this paper, 
we observe that the presence of rising stars is closely correlated with the early diffusion of user interest in social networks, which is validated in the case of Taocode (an intermediary that diffuses user interest in Taobao).  
Thus, we propose a novel framework, RiseNet, to incorporate the user interest diffusion process with the item dynamic features to effectively predict rising stars.
Specifically, we adopt a coupled mechanism to capture the dynamic interplay between items and user interest, and a special designed GNN based framework to quantify user interest. 
Our experimental results on large-scale real-world datasets provided by Taobao demonstrate the effectiveness of our proposed framework. 
\end{abstract}
}

\maketitle

\IEEEdisplaynontitleabstractindextext

\IEEEpeerreviewmaketitle


\section{Introduction}
\label{sec:Introduction}

\centerline{\textit{It's not about what it is, it's about what it can become.}}

\rightline{\textit{---Dr. Seuss} }

In the online market, the item sales change drastically over time. Evaluating whether the item has the potential to increase sales in the future is of great importance in online market. In this paper, we propose to study this novel and practical problem: rising star prediction. We call these promising items \emph{Rising Star} for their ability to rise from low-turnover items to best-sellers in the future. 
For instance, the face mask becomes a rising star at the beginning of 2020, whose sales soared exponentially when COVID-19 began its rapid spread.
The sales of face masks on Taobao reached 80 million from Jan.20 to Jan.21 2020, which is more than one thousand times of that of the last week. 
Although rising stars account for a small proportion of total items and their unstable sales trend adds more difficulties to prediction, the great value brought by them cannot be ignored. 
We outline the importance of rising stars by discussing following common issues in e-commerce. 

\vpara{Alleviate unbalanced recommendation.}The rapid development of recommender systems in e-commerce has significantly helped online-stores to increase conversion rates and sales by matching users with relevant items. 
However, most recommender systems tend to favor a small set of popular items in their recommendations, and disfavor currently unpopular items, even when these items could be potentially preferred by a significant number of users~\cite{liu2019real}. The ''Matthew Effect'' thereby becomes increasingly evident: 
the head items become more and more popular, while long-tail items lacking wide and timely exposure \cite{niu2020dual}\cite{anderson2006long}. 
This issue has caused negative influence on the balance and sustainability of the online market ecosystem,.
In this case, retailers of those long-tail items will turn to other competitors for better opportunities.
Precise identification of rising stars can be used to find out the promising items in long-tail and give them enough exposure, which will balance the recommender system and benefit the whole business environment.

\vpara{Balance supply and demand.}The balance of supply and demand is important for online retailers. 
Untimely adjustment of supply would greatly affect the sales and thus decrease retailers' benefits~\cite{ekambaram2020attention}. 
This issue is more significant for rising stars, as the demand for which usually fluctuates sharply. 
For example, popular and fashionable clothes of the season may have great potential to be a rising star. 
However, if the retailers cannot predict the upcoming increase in sales, users will lost interest as the waiting time for delivery increases~\cite{qi2019deep}. 
On the other hand, if the new item lack the potential to be a rising star, the retailers can prevent over-production in advance by correct prediction.

\vpara{Allocate marketing resources.}In online markets, many promotion campaigns are conducted to stimulate sales~\cite{cummins2010sales}. 
As the high promotion cost and limited marketing resources, it is essential to select the right item with high return benefits. 
Compared with current best-seller, which have low marginal benefits, and other long-tail items without potential, rising stars would more likely to be the first choice of promotion campaign.

Up to this point, we have demonstrated the indispensable role played by rising star prediction in terms of fairness issue (from the perspective of the e-commerce platform), supply-demand issue (from the retailers’ point of view) and marketing strategy. 
Still, finding the solution to the problem remains a great challenge. 
First, compared with other items, rising stars’ sales trend tends to exhibit more suddenness caused by some \textit{external events}. 
For instance, the aforementioned sudden rise in demand of face masks is caused by COVID-19. 
How might we capture such sudden sales changes without any further information about the external event? 
In addition, some rising stars tend to be new arrivals (e.g., iPhone 13) or items that just release from sales suspension; as a result, most of them lack long-term historical sales related data. 
Together with the consideration of external events influence, conventional sales prediction methods can not fit our rising star prediction problem, which mainly adopt time series analysis techniques and consider past long-term sales records. 

To address above challenges, we propose and investigate the interplay between rising stars and the early diffusion of user interest in social networks. 
The idea is inspired by the following two aspects: 
1) External events will be reflected by the diffusion of user interest \cite{weng2011event}\cite{gao2017novel}. 
For instance, COVID-19 led to the large-scale diffusion of user interest in the face mask. 
Therefore, even without the explicit information of external event, we could still perceive it by the diffusion of user interest. 
2) Diffusion of user interest may prefigure future sales information \cite{hollerit2013towards}\cite{zhao2014we}. 
Users that are activated in the diffusion process have a high probability of purchasing the diffused items in the future. This can by explained by
principle of social proof: we tend to purchase items that others have purchased or recommended, especially people who are close or similar to us~\cite{cialdini2007influence}. 

\begin{figure}[t]  
	\centering  
	\includegraphics[width=0.37\textwidth]{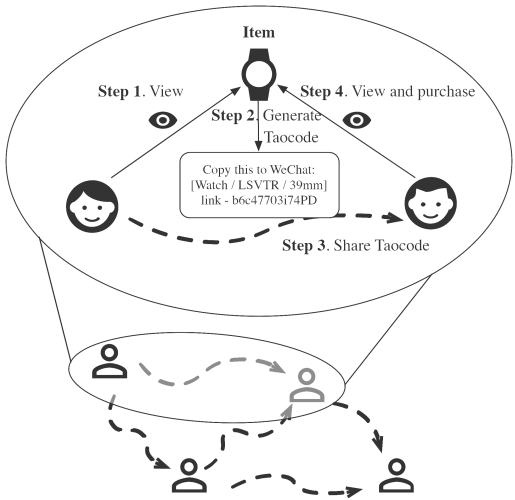}
	\caption{Illustration of Taocode diffusion in social networks. A user can generate the Taocode message of an interested item and share it with specific friends. Afterwards, the receivers are able to further propagate the Taocode to their own friends, and thus creating diffusion. 
    In the diffusion process, any receiver could open the Taocode, view and even purchase the item. }
	\label{fig:intro}     
\end{figure} 

In order to study user interest diffusion, we conduct a case study of Taocode. 
Taocode is used by more than hundreds of millions of users on Taobao to conveniently share items in various online social networks (e.g., Wechat\footnote{The largest multi-purpose messaging application in China.} and Weibo\footnote{A Twttier like social media in China: http://weibo.com}). 
Figure~\ref{fig:intro} demonstrates how Taocodes diffuse in social networks, 
a user generates the Taocode message of an interested item and shares with specific friends. Afterwards, the receivers are able to further share the Taocode to their own friends, and thus the diffusion formed. 
In the diffusion process, any receiver could open the Taocode, view the corresponding item, and even purchase it.  

To conduct our study, we employ a real-world dataset, which contains 368 million records of Taocode diffusion, and relevant to 18 million items and 72 million users. 
In addition, we have 204 million purchase records.  
Our empirical observations show clear  
mutual interactions between Taocode diffusion and item purchase data:  
For instance, we find that diffusion scale (i.e., the number of users who receive the message) 
positively correlates with sales. 
Moreover, the sale volume of items propagated by ``hub users'' are more than five times of that of those haven't been propagated by ''hub users''. 
More importantly, we observe that rising stars tend to exhibit greater fluctuation in their diffusion statistics (e.g., diffusion scale), and they are more likely to be propagated by more ''hub users'' compared with other items. Also, we illustrate the importance of dynamic information in rising star prediction.

Despite of interesting insights provided by our observations, 
the question of how to jointly model the likelihood of an item to become a rising star and the dynamic Taocode diffusion into a unified framework is challenging. One straightforward method involves concatenating item historical data and statistical data of Taocode diffusion at each temporal point to capture the overall patterns. However, only using statistical data of Taocode diffusion will lose the rich information at user level. As Taocode diffusion data could naturally be regarded as dynamic graph, we propose a graph neural network (GNN) based model to quantify user interest. Unlike traditional GNN, we differentiate Taocode influence to users with consideration of both user item preference and Social Proof; and we also differentiate the roles played by different users in the diffusion graph to help graph information extraction. The item temporal data (such as sales) is essential in our task, the simple concatenation between the item and Taocode sources precludes capturing the significant mutual influences between them (Section~\ref{sec:observation}). Here, we highlight our coupled mechanism in \model{}, which combines information from both sides while modeling the interplay between them at each time step. As dynamic nature of our data, \model{} further models the dynamic changes at each time step and adopts a multitask mechanism to better leverage the information.
\hide{
2) a multi-task learning framework to quantify the user interest, including a self-supervised task on dynamic graphs to auxiliary\yx{this word ok?} rising star prediction task on our designed graph neural network, which is suffered \yx{too severe?}from class-imbalance and the consequent sparse information problem.
\yang{unclear, and too many details}
3) a hierarchical dynamic structure that models the temporal dependency of the unified information and predicts the rising stars in an end-to-end manner. 
}
Experiment results on four Taobao real-world datasets demonstrate the effectiveness of our model \model{} by achieving significant predicting score against existing alternatives.

Accordingly, the major contributions of our work can be summarized as follows: 
\begin{itemize}[leftmargin=*]
    \item We propose to study the novel and practical problem: rising star prediction.
    \item Based on observational studies, we formulate a novel framework \model{} to predict rising stars by leveraging user interest diffusion and item information simultaneously. 
    \item Employing a large real-world dataset, we conduct extensive experiments to validate the superiority of our proposed model against several baselines. 
\end{itemize}

\section{Preliminaries}
\label{sec:Preliminaries}


\begin{figure*}[t]
	\subfigure[]{
	\includegraphics[width=0.23\textwidth, height= 80pt]{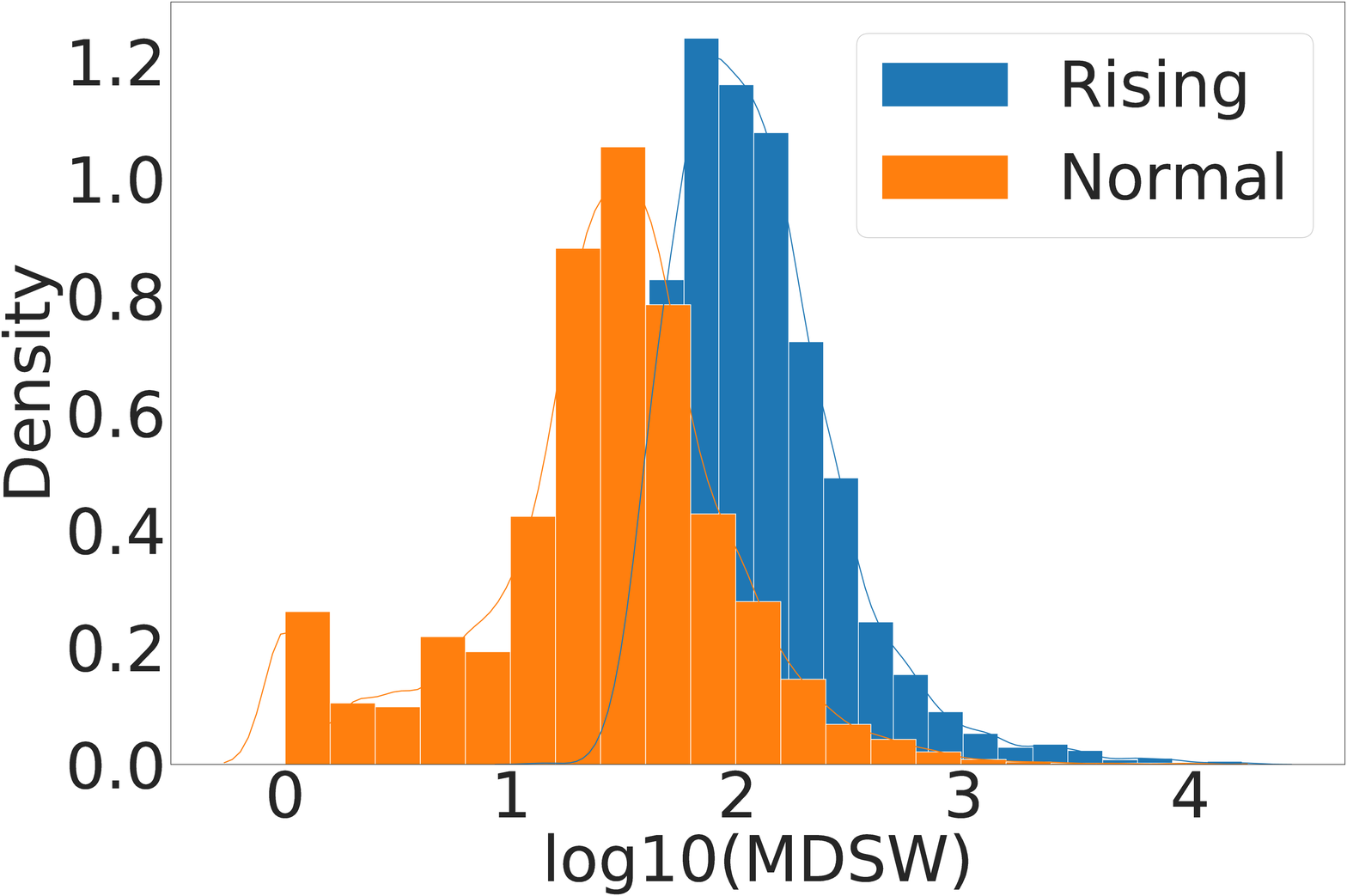}
	\label{3}
	}
	\subfigure[]{
	\includegraphics[width=0.23\textwidth, height=80pt]{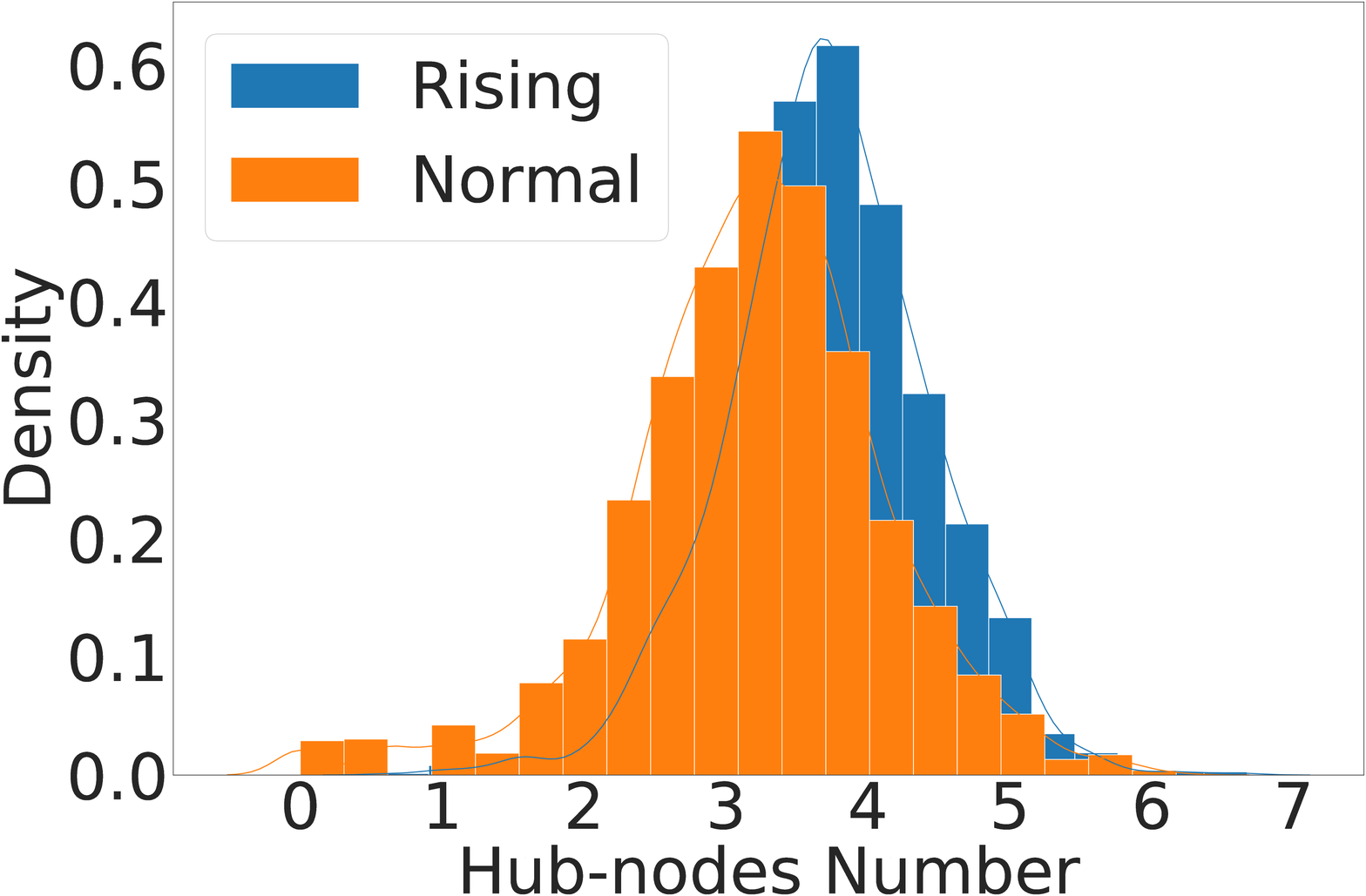}
	\label{4}
	}
	\subfigure[]{
	\includegraphics[width=0.23\textwidth, height=80pt]{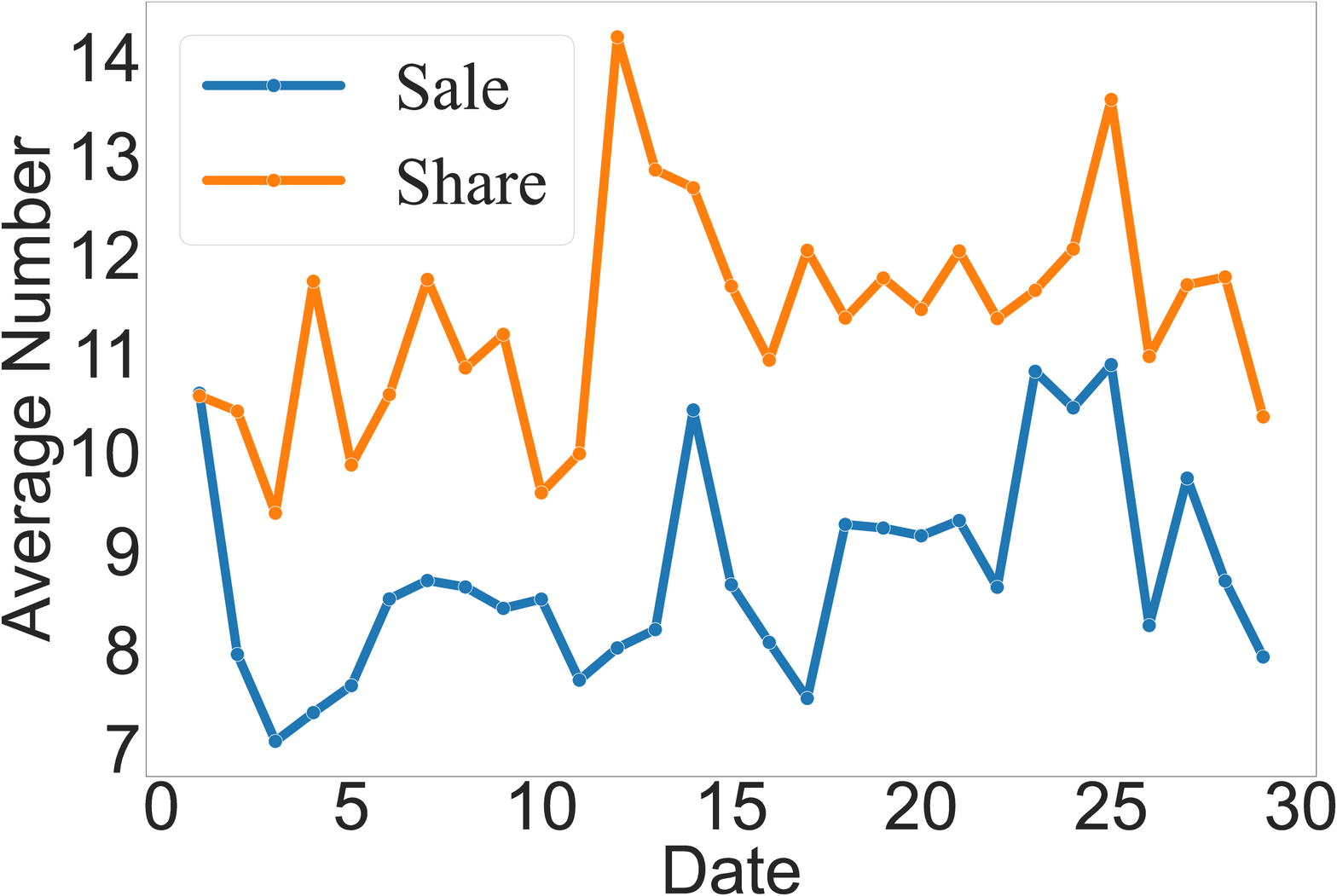}
	\label{1}
	}
	\subfigure[]{
	\includegraphics[width=0.23\textwidth, height=80pt]{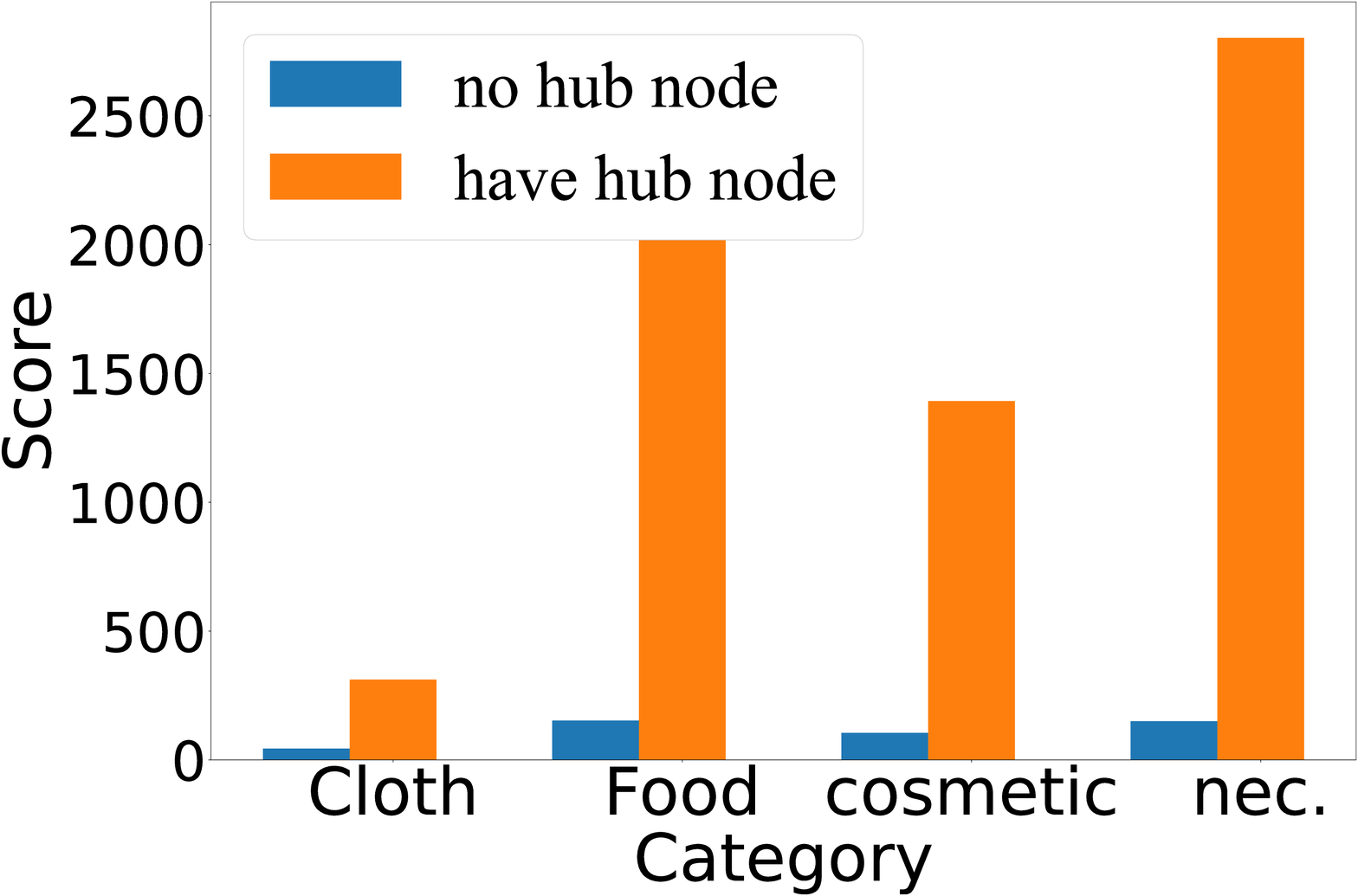}
		\label{2}
	}

    \caption{Observation results of the Taocode diffusion network. (a) The distribution of maximum difference between the graph scales during two consecutive weeks (MDSW) of rising stars and normal items. (b) The diffusion graph hub node numbers' distribution of rising stars and normal items. (c) Interaction between item sales and diffusion graph scales. (d) Relationship between item sales and diffusion graph hub node numbers. }
\label{fig:exp:params}
\end{figure*}


\hide{
\subsection{Notations}{
	In this paper, we use lower-case letters to indicate scalar parameters and bold letters for vectors; moreover, the superscript $i$ of the bold vector symbol represents the $i^{th}$ dimension of the vector, with the subscript $u$ indicating that the bold vector is for vertex $u$. More details are presented in the following table. 
	\makeatletter\def\@captype{table}\makeatother
	\caption{Notations.}
	\label{tab:freq}
	\begin{tabular}{ccl}
		\toprule
		Symbols & Meanings \\
		\midrule
		$V$ & Set of vertices in the given network\\
		\midrule
		$E$ & Set of edges in the given network\\
		\midrule
		\bm{$i$} & product \\
		\midrule
	\end{tabular}
}
}
\subsection{Datasets}

Our dataset comprises three sets of real-world data, all of which are provided by China's largest e-shopping platform, Taobao. 
We construct our dataset by sampling records during a time period of three continuous months in 2020. 


\vpara{Taocode diffusion data}
This dataset contains 368,888,906 records of Taocode diffusion among 72,594,418 users, involving 18,325,582 items. 
Taocode is a special form of sharing items among Taobao users which we have introduced in Section~\ref{sec:Introduction}. 
We define a Taocode diffusion record for the shared item $i$ as a 4-tuple $e$ = ($i$, $u_s$, $u_r$, $t$), where $u_s$ denotes the Taocode sender, $u_r$ denotes the Taocode receiver, and $t$ is the timestamp when the record takes place.   

\vpara{Taobao purchase data}
This dataset covers 204,398,128 online purchase records on Taobao, involving 215,821,800 users and 116,867,875 items. 
Each purchase record contains the information of the buyer, the item, the turnover, and the purchase time. 

\vpara{Basic information data of Taobao items}
This dataset covers basic information of 31,093,128 items, such as name, price, and category in a three-level pattern. 
For example, basic information of an ice cream is recorded in the data as (Hagen-Dazs, 49, Food, Frozen food, Ice cream). 
\hide{
\begin{table}
    \caption{Description of some major notations}
	\begin{tabular}{cc}
        \hline
		Notation&Description\\
		\hline
		$i$& The item on the online retail platform\\
		$u$& The user on the online retail platform\\
		$e$& The Taocode record on the Taobao platform\\
		\hline
		
	\end{tabular}
	\label{tab:notation}
\end{table}
}
\subsection{Definitions}
\label{sec:data:prob}
In this section, we provide necessary related definitions.
For notations, we use lower-case letters to denote scalars (e.g., $k$) or single object (e.g., $i$), and upper-case letters for sets (e.g., $E$).

\vpara{Rising Star.} 
 \emph{We define items that rise from low-turnover items to best sellers as rising stars.}
To suit the demand of the application scenario of Taobao, we define regard an item that is not ranked within the top 3\textperthousand~of its high-level category in the first four weeks, but later rises to be ranked in the top 1\textperthousand~over the next two weeks as a rising star. 
Items are ranked in their own high-level categories. It is because there are differences between the average sale volume of different categories. For example, the average sales of items in food is more than three times higher than that in clothing. 

\par

\vpara{Dynamic Taocode Diffusion Graph.} 
 For an item $i$, dynamic Taocode diffusion graph of item $i$ is a sequence of Taocode diffusion graphs $G_i = \{G_i^0, G_i^1, G_i^2,..., G_i^T\}$. We refer the graph constructed by item $i$'s Taocode diffusion records at the time step $t$ as the Taocode diffusion graph $G_i^{t}$. The Taocode diffusion graph $G_i^{t}$ = ($V_i^{t}$, $E_i^{t}$, $M_i^{t}$), where $E_i^{t}$ denotes the set of Taocode diffusion edge, $V_i^{t}$ denotes the set of users (nodes), and $M_i^{t}$ denote the user feature matrix.
\par

\vpara{Rising Star Prediction.}
In light of above notations, we can define the rising star prediction addressed in this paper as follows: given a specific item $i$ on the online platform, the goal is to estimate $P(y_i|G_i,X_i)$, which is the probability that item $i$ will become a rising star ($y_i=1$) or not ($y_i=0$). $X_i$ denotes item $i$'s temporal features, which can be represented as a sequence $X_i = \{X_i^0, X_i^1, ..., X_i^T\}$.




\section{Exploratory Analysis}

\label{sec:observation}

\hide{
\begin{figure*}[ht]
	
	\centeringb
	\includegraphics[width=1\textwidth]{./fig/ob1.PNG}
	\caption{Observation results of Taocode diffusion network. (a) interaction between item sales and diffusion graph scales. (b) relationship of item sales and diffusion graph's hub node numbers. (c) the maximum weekly gradient of scales' distribution of rising stars and other items. (e) the diffusion graph contained hub node numbers' distribution of rising stars and other items.}
	
\end{figure*}}
In this section, we first study the interplay between user interest diffusion (Taocode diffusion) and rising star identification, and then observe the interplay between user interest diffusion and item features. Furthermore, we examine whether dynamic information is needed in rising star prediction.
\subsection{Observing Rising Stars Through Taocode Diffusion}
\label{sec:uvr}
In order to study the relationship between rising star identification and Taocode diffusion graph, we respectively sample 2500 rising stars from clothing category during a period of specific six weeks in 2020, and 2500 normal items to  conduct  the  observation  experiments. Here, items that do not rise from low-turnover items to best sellers are defined as normal items. To comprehensively observe the correlation, we propose to investigate the Taocode diffusion graph from two different levels: graph level and node level.

Firstly, on the graph level, we aim to study whether the rising stars exhibit patterns that different from the normal items on diffusion graph. To examine this, we draw the distribution of the maximum graph scale difference between two consecutive weeks (MDSW for short) for the group of rising stars and the group of normal items during the specific six weeks.
\begin{equation}
    MDSW=Max(s_i^m-s_i^n)_{1<=n<=5;m=n+1}
\end{equation}
where $s_i^m$ denotes the sales volume of item $i$ in week $m$.

In Figure 2(a), there is an obvious rightwards deviation of MDSW distribution of rising stars compared with that of normal items, which indicates the rising stars exhibit greater fluctuation in their diffusion graphs. This result shows that the Taocode diffusion graph can provide us with information to identify rising stars from the graph level.

Furthermore, we want to study whether the diffusion graph can provide information to identify the rising star from the node level. Here, we take the hub nodes as an entry point to examine this question.
In the Taocode diffusion graph, hub nodes represent the most active users. In our experiments, we first quantify hub nodes as the users who have the top 1\% rank number of Taocode sharing records in the Taocode diffusion data.
We draw the distribution of the numbers of hub nodes that contained in each sampled rising stars' diffusion graph during the same specific six weeks 2020, and we draw the distribution of the normal items group for comparison. The distribution in Figure 2(b) illustrates that items whose diffusion graphs contain more hub nodes are more likely to be rising stars. From the subsequent case study, we find the reason behind this result. The hub nodes are the most active users on Taobao e-commerce platform, who have greater abilities to identify potential items depending on their abundant online shopping experience. Moreover, certain hub nodes (such as internet celebrities) can employ their significant online influence to recommend a particular item and thus create a rising star. To this end, we find that diffusion graph is also helpful from the node level in rising star identification.

Together, from above observations, we get the conclusion that the Taocode diffusion graph has close correlation with rising star identification from both graph level and node level.

\subsection{User Interest Diffusion Vs. Item Features}
\label{sec:uvi}

As stated in Section~\ref{sec:Introduction}, user interest diffusion is naturally related with item features. To further study this relationship, we also conduct observations on Taocode diffusion data from graph level and node level respectively. 

First, from the graph level, we observe whether diffusion graph statistics (such as diffusion scale) correlate with item features (such as sales). We sample one million items from clothing category in our dataset and then examine their average sales and average diffusion graph scales that changes over one month in 2020. 
From the results shown in Figure 2(c), we can see that the item sales is directly proportional to the scale of Taocode diffusion graph, which validates the close correlation between item sales and user interest. 
This result is consistent with our recognition in real-world situations: item sales directly guide the interest of users due to the herd mentality~\cite{foxall2002consumer} (users tend to pay more attention to items with high sales numbers because they think these items are more reliable); meanwhile, user interest diffusion greatly influences the item sales (for example, active sharing among users via Taocode can greatly promote item sales). 

As we have observed that the scale of diffusion graph dynamically interact with item sales from the graph level, we furthermore study the relationship from the node level. Here, we also conduct observations on hub nodes as an example. Owing to the sales gap between different categories which we have discussed in Section~\ref{sec:Preliminaries}, we respectively sample one million items from clothing, food, cosmetic and necessities categories, and then label hub nodes according to Taocode sharing ranks of each category. 
We divide the sampled items into two groups depending on whether or not they were recommended by hub nodes within the selected one month. We compare average sales of each group's containing items from four categories; results are shown in Figure 2(d). The items that have been recommended by hub-nodes are found to have far higher sales in all four categories. The reason is that hub nodes are typically influential users, such as, popular bloggers, who are more likely to diffuse selected items' Taocodes to their followers, and professional buyers who actively send Taocodes to their customers. Users are not only more likely to receive Taocodes from these influential users, but also more willing to take their advice and then purchase the recommended items online. From the node level, we can see that Taocode diffusion graph also has its correlation with item sales.

From both the graph level and node level, we find close relationship between item features and user interest diffusion. Next, we want to explore whether the dynamic information is necessary in rising star prediction.



\subsection{Importance of Dynamic Information}

\label{sec:ob-dynamic}

\begin{figure}[h]
	\subfigure[]{
	\includegraphics[width=0.23\textwidth, height=80pt]{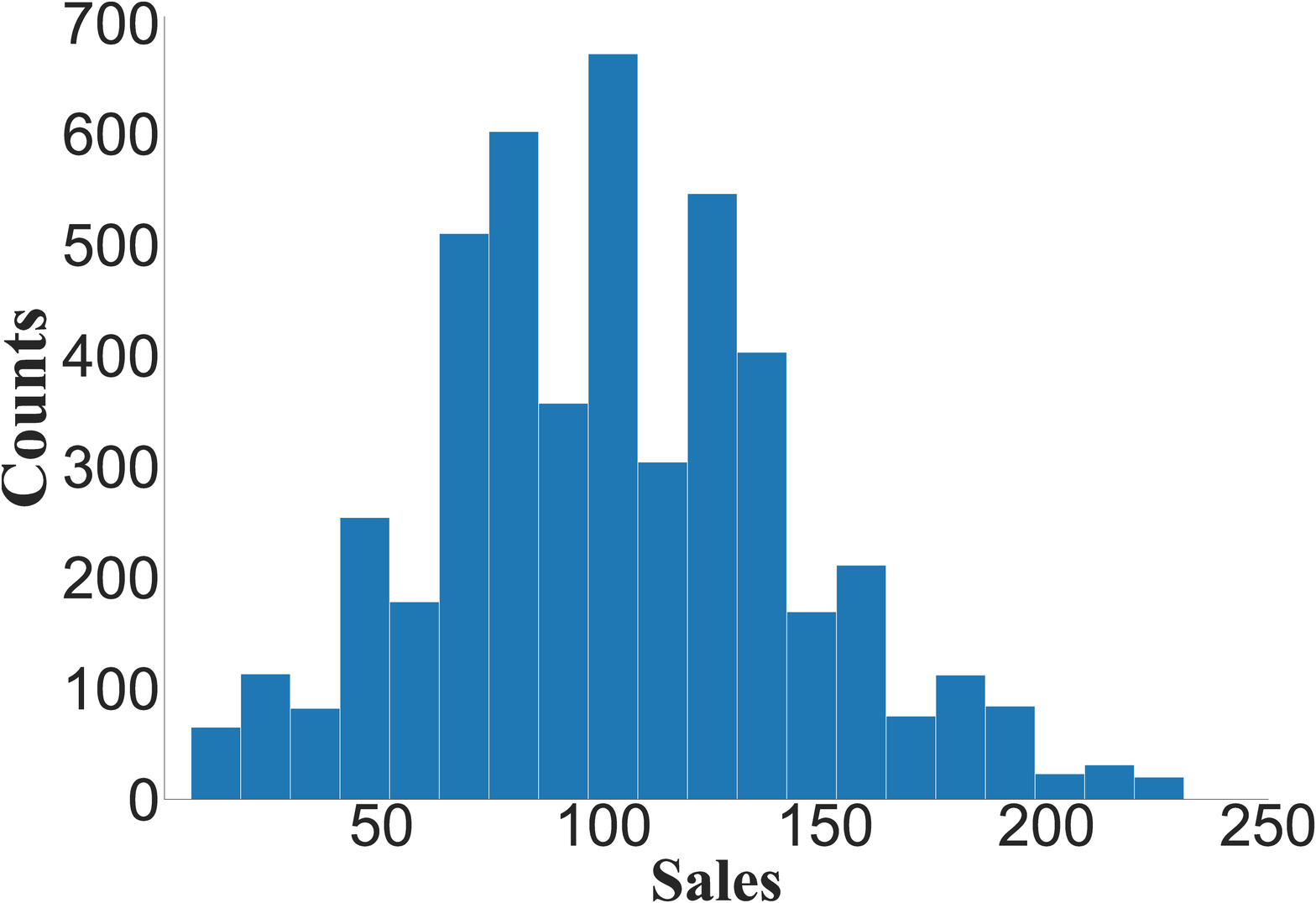}
	\label{1}
	}
	\subfigure[]{
	\includegraphics[width=0.23\textwidth, height=80pt]{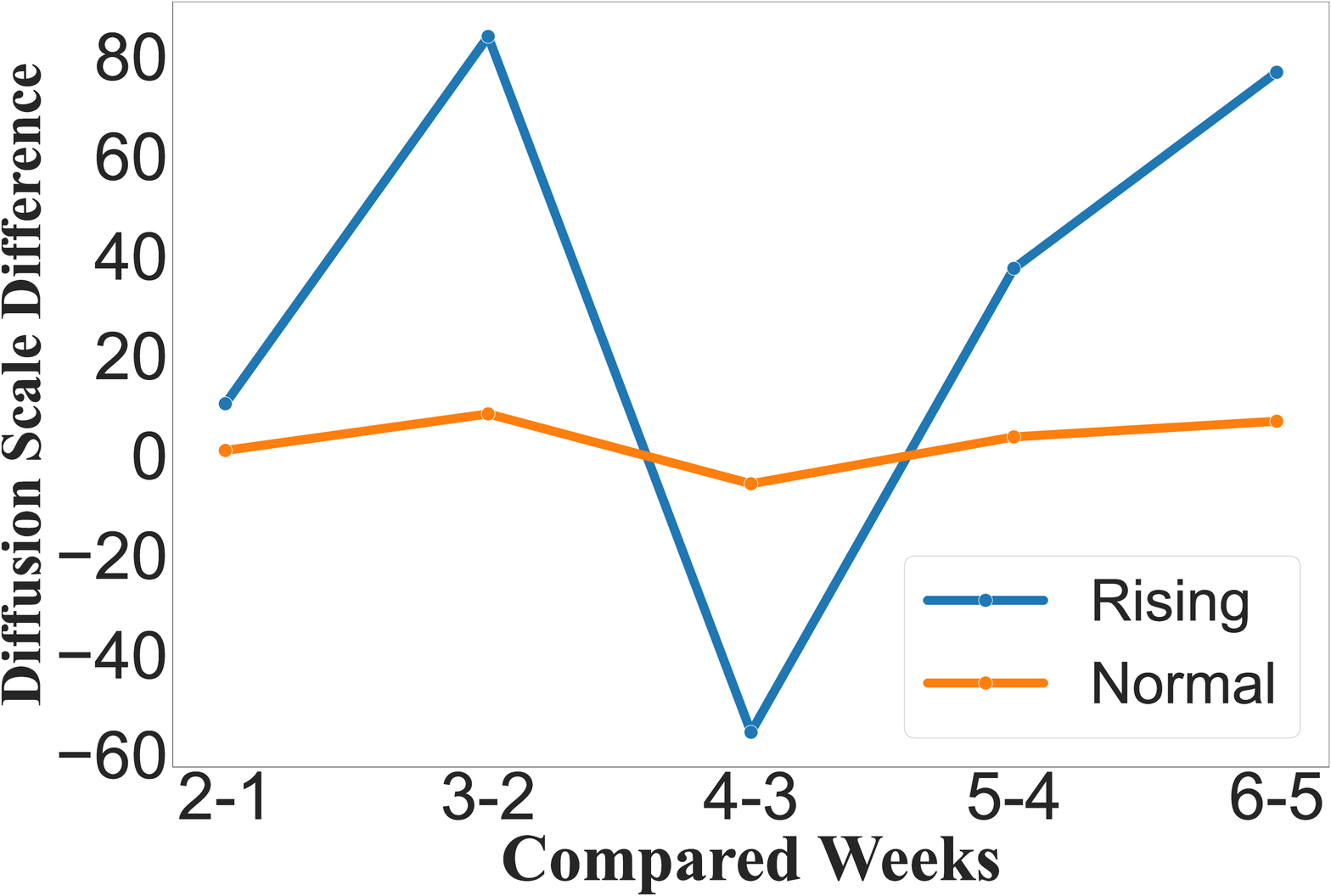}
		\label{2}
	}
	
    \caption{Observation of the dynamic information. (a) The next week's sales diffusion of the items with the same sales the last week. (b) The average weekly diffusion scale difference over the last six weeks for rising star and normal items that have the same diffusion scale over the last week.}
\label{fig:ob_dy}
\end{figure}

In online markets, the item information is naturally dynamic. It drastically changes with time, which includes sales, price, etc.
What if we ignore this dynamic information and simply predict item sales with the information from last time step?   
To answer this question, from items in clothing category, we sample all items which have the same sales volume as average in a specific week.
Then we observe the sales volume distribution of these items in the next week, shown in the Figure~\ref{fig:ob_dy}, which is not concentrated. 
This result suggests that items with same information in the last time step is much more likely to show divergent sales trend in the future. 
In other word, more historical information is necessary to make the prediction more accurate.
 
Like item features, user diffusion graph also evolves with time. 
Here, we investigate whether the dynamic information in diffusion graph can further help us to identify rising stars when the information from the last time step is not the only condition. 
We sample all items that have the same diffusion graph scale as average in a specific week, and divide them into two groups according to whether they are rising stars in the next week. 
After that, we draw the average weekly diffusion scale difference (fluctuation) of each group with historical information in the last six weeks. 
As presented in Figure~\ref{fig:ob_dy}, although both groups have the same diffusion scale in the last week, rising star group shows more volatility in the diffusion scale difference in the last six weeks compared with the other group. 
This result shows that with the help of historical diffusion information, we can identify rising stars that cannot be identified with only condition of information from the last time step.

To this end, we illustrate the importance of utilizing the dynamic information in rising star prediction.

\section{Our Approach} 
\label{sec:model}

\begin{figure*}[htp]
	\centering
	\includegraphics[width=18cm]{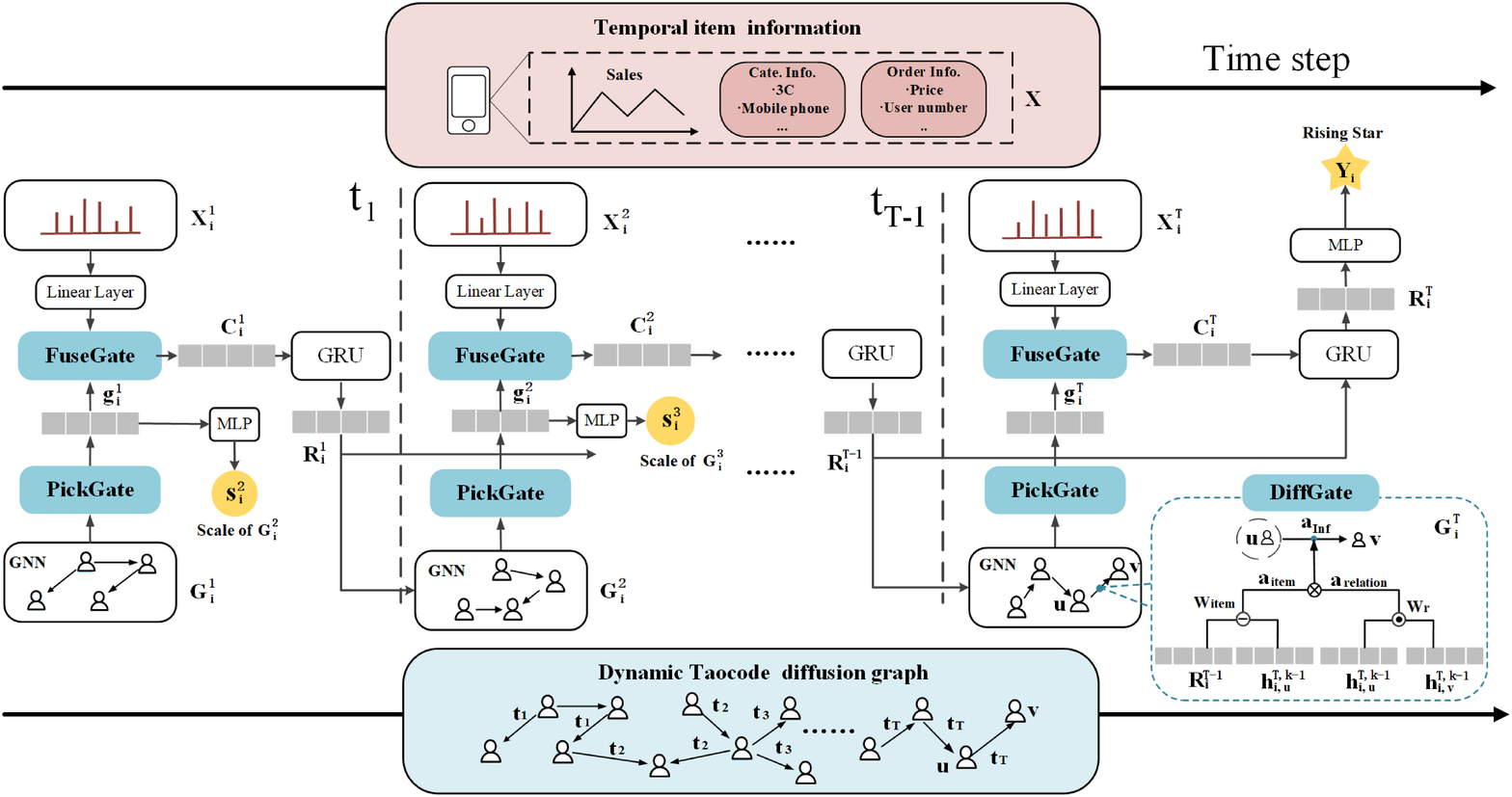}
    \caption{The structure of \model{}. Given dynamic Taocode diffusion graph and item temporal information, \model{} captures the interplay between them through a coupled mechanism: a DiffGate (model various influence of Taocodes on user interest), a FuseGate (unify item and use interest information). A PickGate in GNN is to distinguish the roles of different nodes when generate the graph representation with node embeddings. A multitask framework is used in training to better leverage user interest in the dynamic diffusion graphs. The GRU cells are responsible for modeling temporal dependency.}
	\label{fig:ModelArch}
\end{figure*}                                                                             
In this section, with inspirations derived from Section \ref{sec:observation}, we propose our model, \model{}, with the general idea to incorporate user interest diffusion with item information to predict rising stars. 
From the observation results in Section~\ref{sec:observation}, we validate the following conclusions:
\begin{enumerate}
\item User diffusion graphs can provide great support for rising star prediction task from both graph-level and node-level.
\item There is a strong mutual interaction between the user interest diffusion and item features.
\item The dynamic information contained in item features and diffusion graph is helpful for rising star identification.
\end{enumerate}

Inspired by above observation results, we propose the \textit{\model{}} model.
\subsection{General description}
Here, we elaborate the general framework of \model{}. 
As illustrated in Figure \ref{fig:ModelArch}, \model{} aims to incorporate both the diffusion process of user interest and the dynamics of item features to effectively predict rising stars.

As the diffusion process could naturally be regarded as dynamic graphs, we propose a graph neural network based framework to fully capture the information in Taocode diffusion from both graph and node level. Unlike the traditional GNN models, we differentiate the roles of different users when we generate the graph representation and model each Taocode diffusion influence in the propagation process with consideration of both user item preferences and \textit{social proof}.
Meanwhile, considering the importance of item temporal information, we also model the item information from item temporal features (such as sales and price). Then we combine the item and user interest information to get a unified representation while capturing mutual interactions between them at each time step. Considering the dynamic nature of our data, we further build the temporal dependency of the unified representation at each time step, capturing the dynamic changes to help identify rising stars.

For details, the core components of \model{} are elaborated in the remaining of this section.

\subsection{Modeling User Interest Diffusion Graph}
\label{sec:user}
To quantify the dynamic user interest information in Taocode diffusion graph, here we introduce our designed graph neural network based framework. Naturally, GNNs can be applied to quantify the user interest due to their ability to model the non-linear correlation of the graph structure and user interactions in diffusion. Also, in the GNN user propagation process, the model could capture the collaborative filtering signal and thus strength user interest learning and improve predicting effectiveness~\cite{DBLP:conf/sigir/JinG0JL20}. 

More specifically, we introduce the overall pattern of our GNN module. In the GNN propagation process, as different Taocodes affect user interest to varying degrees, it is intuitive that different Taocodes (edges) should be assigned to different weights. DiffGate models the various Taocodes' influence to users by considering both user item preference and user relationship influence. After propagation, we then generate the Taocode graph representation with the learnt node embeddings. In the diffusion graph, a small part of users (hub nodes) often play decisive roles in the overall user interest change (Section~\ref{sec:uvi}). Our designed PickGate picks these hub nodes and conducts the readout function on the hub nodes and all nodes respectively. Then we get the final graph representation by concatenating them. PickGate could avoid the important information contained in hub nodes to be over-smoothed by other nodes, which helps to retain more user interest information in the graph representation. In the GNN training process, multitask mechanism is adopted on GNN module, which helps GNN to converge the user interest information from dynamic Taocode graphs. 

The details of GNN module are introduced in the following:

\vpara{DiffGate.}
DiffGate models the various influence of different Taocodes to users, which helps to better capture user interest in the propagation process. 

Generally speaking, a Taocode's influence on user interest is mainly affected by the following two factors: 1) dynamic item information. When Taocodes share different items, the influence they bring to users is naturally different. A user tends to pay more attention to the Taocodes which share his interested categories' items. For example, compared to Taocodes that share rock music CDs, those share piano CDs will attract classical music lovers more easily. Moreover, even with the same item, the observation results in Section~\ref{sec:uvi} show that the user interest will be greatly influenced by the item's current features (such as sales and price);
2) relationship between the Taocode sender and receiver. According to the principle of social proof ~\cite{cialdini2007influence}, user interest will change to varying degrees according to different relationships between the Taocode sender and the receiver. People tend to accept the Taocode recommendation from their close friends. Also, some special types of relationships, such as teacher-student relationship, will greatly affect the user’s interest towards the Taocode.

Simultaneously considering the above two factors of influence, Diffusion gate (DiffGate) models the probability that a Taocode will influence a user's interest to help the GNN capture the user interest information in the propagation process. More specifically, we first calculate two weights respectively: item attention for modeling dynamic item information's influence; and relationship attention for modeling user relationship' influence. And then we unify these two weights to get the final user interest influence possibility of a Taocode. 
For a Taocode, which shares item $i$ between the user $u$ and the user $v$, we define its item attention for the user $u$ as following,
\begin{equation}
	a_{u,v,i^{item}}=W_{item} \cdot [h_{i,u}^{t, k-1} - I_i^{t-1}]
\end{equation}
where $h_{i,u}^{t, k-1}$ represents the user $u$'s hidden embedding in the $G_i^{t}$ diffusion graph and layer $k-1$ of the GNN model. $I_i^{t-1}$ is the representation of item $i$, which is generated at the last time step; this will be introduced in Section ~\ref{sec:couple}. $W_{item}$ is a learnable vector to project the embedding to the influence probability. Note that we don't directly concatenate the item and user embeddings to do the projection. The reason is such attention could only get the global importance, as discussed in \cite{ShakedBrody2021HowAA}, and the ranking of the item attention scores is unconditioned on the query user, which is inconsistent with our intention to capture the various user interest towards Taocodes of different items.

For the relationship attention of a Taocode diffusion, we define it as:
\begin{equation}
	a_{u,v,i^{r}}=W_{r} \cdot [h_{i,u}^{t, k-1} \odot h_{i,v}^{t, k-1}]
\end{equation}
where $W_{r}$ is a also learnable projection vector.

With the calculated item attention and relationship attention, we combine these two attentions and calculate the Taocode influence probability $a_{u,v,i}$ for user $u$ as:
\begin{equation}
	a_{u,v,i}=DiffGate(h_{i,u}^{t, k-1}, h_{i,v}^{t, k -1},I_i^{t-1})
\end{equation}  
\begin{equation}
	DiffGate(h_{i,u}^{t, k-1}, h_{i,v}^{t, k -1},I_i^{t-1}) = a_{u,v,i^{item}} \cdot a_{u,v,i^{r}}
\end{equation}  

With the calculated Taocode influence probability $a_{u,v,i}$, for each user $u$ in a given user diffusion graph $G_i^{t}$, we first conduct the aggregation in each GNN layer:
\begin{equation}
	h_{i,u}^{t, k} = h_{i,u}^{t, k - 1} + \sum_{v \in N(u)}a_{u,v} \cdot h_{i,v}^{t, k - 1}
\end{equation}
Here, $N(u)$ denotes the set of neighbor nodes of $u$. The initial user representation is from $M_i^t$, the user feature matrix.

\vpara{PickGate.}
After a fixed number $L$ of layers of GNN propagation, We want to generate the Taocode graph representation with the learnt node embeddings. PickGate is designed here to retain more important information in the final graph representation to help rising star prediction.
As discussed in our observations, there are some important users (named hub nodes) in the diffusion graph. They may be the internet celebrities or professional buyers, even the seller of the item. These hub nodes share a large amount of Taocodes and take the decisive role in the evolving of diffusion graph. Thus, instead of directly applying a readout function on all user embeddings to get the graph representation, PickGate picks these hub nodes in advance. Then we respectively apply the readout function to their embeddings and all nodes' embeddings respectively. And we get the final graph representation by concatenating them. The PickGate could avoid the important information that contained in hub nodes to be over-smoothed by other users. In the PickGate, we adapt a learnable vector to project the representation of each user to get their importance score~\cite{DBLP:conf/icml/GaoJ19}. According to the importance score, we pick out top K\% users,
\begin{equation}
	idx = PickGate(\{h_{i,v}^L, v \in V \},\lceil KS\rceil)
\end{equation}
where $S$ is the number of nodes. 

Then we get entire diffusion graph representation $g_i^{t}$, 
\begin{equation}
\begin{aligned}
    g_i^{t} = &READOUT(\{h_{i,v}^L, v \in V \}) ||\\               
    &READOUT(\{h_{i,v}^L, v \in V(idx) \})
\end{aligned}
\end{equation}
Here, READOUT(*) is a summation function. We prefer the summation function to the average function, as we want to retain more global structural information. 


\vpara{Multitask mechanism.}
Compared with the static graph, the structure of dynamic graph changes drastically over time. In the GNN training process, the drastic changes on the dynamic diffusion graph make the amount of information provided by a single supervisory task insufficient for our task. 
What makes the task more challenging is that our target label is unbalanced due to the scarcity of rising stars.
A designed multi-task mechanism helps the GNN model to capture the user interest more accurately while solving the problem of insufficient supervisory information.

Specifically, in addition to the global supervision task (rising star prediction) on dynamic graph at all time steps, we add a local supervision task (diffusion scale prediction) at each time step to help the GNN to better learn the user interest information.
The diffusion scale prediction task can help the model to recognize the rising stars from the following aspects: 1) From the observation results in Section~\ref{sec:uvr}, the scale information of the diffusion graph can be used to distinguish rising stars. 2) The task can help the GNN to understand the user sharing and forwarding behaviors, which is helpful for the model to understand the user interest towards items. 

More specifically, in each time step $t$ $\in$ ${1,...,T-1}$, we predict the scale $s_i^{t+1}$ of the next time step's diffusion graph ${G_i^{t+1}}$. In this way, the scale prediction task helps us to find out the rising stars in a joint-training fashion with the target task. The predicted scale $\hat{s}_i^{t+1}$ is obtained via a multi layer perceptron: 
\begin{equation}
	\hat{s}_i^{t+1} = W_{s2} \cdot ReLU(W_{s1} \cdot g_i^{t} + b_{s1}) + b_{s2}
\end{equation}
$Loss_0$ is defined as a mean relative square error (MRSE) loss, which is a common evaluation metric in scale prediction task:
\begin{equation}
	Loss_0 =\frac{1}{N}\sum_{t=1}^{T-1}\sum_{n=1}^{N}(\frac{\hat{s}_i^{t+1}-s_i^{t+1}}{s_i^{t+1}})^{2}
\end{equation}
Here, N is the total number of dynamic diffusion graphs (batch size), while T is the total number of time steps.

\subsection{Coupling Dynamic User Interest with Temporal Item Information}
\label{sec:couple}
In addition to the user interest information contained in Taocode diffusion graph, the item's historical features are naturally important for the rising star prediction task. Thus, we aim to combine the information from both parties to predict the rising star accurately.
As we discussed in Section \ref{sec:observation}, the item information and user diffusion information influence each other dynamically. Capturing this interplay is the key to enhance our model's ability in rising star prediction. Therefore, we apply a couple mechanism to combine user interest and item information while capturing this sophisticated correlation between them.
This mechanism comprises two main parts: \emph{DiffGate} and \emph{FuseGate}. In the coupled mechanism, firstly in each time step, the FuseGate combines the user interest and item temporal features' information while also capturing their mutual dependency to generate the item representation. And in the next time step, the DiffGate incorporates the generated item representation to calculate the item attention which helps to capture the user interest. Then the captured user interest information will be used as the input into the FuseGate at the next time step. In this way, a dynamic process of mutual influence is formed by our couple mechanism.

\vpara{FuseGate.}
Once we have obtained the graph representation $g_i^{t}$ generated by GNN at time step $t$, the next challenge we face is that of how to unify the graph information with the input of the item temporal features $X_i^t$. A fuse gate (FuseGate) combines the information from these two parties while also capturing their mutual dependency. In the fuse gate, we first use a learnable vector to get the item temporal representation from the initial features of the item. Then, the fuse gate models the mutual dependency of user interest and item information through a multi layer perceptron (MLP). After the MLP, the calculated vector is passed through a soft-max function, and then multiplied the concatenated graph representation and item representation to yield the representation of the combined information $C_i^t$:

\begin{equation}
	C_i^t =FuseGate(g_i^t,X_i^t)
\end{equation}
\begin{equation}
\begin{aligned}
	FuseGate=&SoftMax(MLP(g_i^t||W\cdot(X_i^t)))\\
	        &*(g_i^t||W\cdot(X_i^t))
\end{aligned}
\end{equation}

$C_i^t$ denotes the unified representation of item $i$ at time step $t$. which is then used by the dynamic model presented in Section~\ref{sec:dynamic} to generate item current presentation $I_i^{t}$. Item current presentation $I_i^{t}$ will be used in the next time step's DiffGate. $g_i^{t}$ is the generated graph representation with the help of DiffGate.

\subsection{Capturing the Temporal Dependency}
\label{sec:dynamic}

Our data is naturally dynamic. After exacting and unifying the information at each time step, and considering the importance of dynamic information (discussed in Section~\ref{sec:ob-dynamic}), a time series model allows us to flexibly utilize the temporal unified information $C_i^t$ and build the temporal dependency of dynamic data among the time steps. Here, we take a recurrent model with GRU cells~\cite{cho2014learning} as an example; at time step $t$:
\begin{equation}
	z^{t} = \sigma(W_zC^t + U_zI^{t-1})
\end{equation}
\begin{equation}
	r^{t} = \sigma(W_rC^t + U_rI^{t-1})
\end{equation}
\begin{equation}
	\hat{I}^{t} = tanh(WC^t + U(r^t\odot I^{t-1}))
\end{equation}
\begin{equation}
	I^{t} = (1 - z^t)\odot I^{t-1} + z^t\odot\hat{I}^{t}
\end{equation}

\noindent where $\odot$ is an element-wise dot operation. We share the GRU parameters for all the items and all the time steps. The initial input $I_i^0$ is a vector of all ones. Use of the GRU cell allows a flexible combination of previous item representations $I_i^{t-1}$ and the newly arrived unified input ${C_i^t}$ at time step $t$. GRU adopts two gates: the update gate $z$ and the forget gate $r$. The forget gate $r$ controls how much of the previous item representations $I_i^{t-1}$ should be forgotten, while the update gate $z$ controls how the newly arrived concatenated input ${C_i^t}$ and the processed previous item representations $I_i^{t-1}$ are combined.

After $T$ time steps, by means of the iterative interplay between GNN and GRU cells, we obtain our final representation $I_i^{T}$ for an item $i$. Then, we transform the representation $I_i^{T}$ via a multi layer perceptron and a softmax function to get the predicted result $p_i$, 
\begin{equation}
	\hat{y}_i = Softmax(ReLU(W_2 \cdot ReLU(W_1 \cdot I_i^{T} + b_1) + b_2))
\end{equation}
We define the rising star prediction loss as a cross-entropy loss:
\begin{equation}
	Loss_1 = -[y_ilog(\hat{y}_i)+(1-y_i)log(1-\hat{y}_i)]
\end{equation}
where $y_i$ is the true label of the rising star prediction task. 

Note that $Loss_0$, which is introduced in Section~\ref{sec:user}, is only used to update the GNN's parameters in a joint-training fashion with $Loss_1$. The loss of the graph neural network is defined as:
\begin{equation}
	Loss_g = Loss_0+Loss_1
\end{equation}
In this way, the graph neural network are trained jointly by the prediction of rising star and the next time step's scale prediction task. 
The multi-task learning further improves the generalize ability of our model.

\section{Experiments}

\begin{table}
    \caption{Overview of experimental datasets \\(R. s. rate represents Rising star rate; sale in. means sales increase multiple from week one to week six)}
    \resizebox{88mm}{9mm}{
	\begin{tabular}{c|ccccc}
        \hline
		Dataset&User&Taocode&Purchase&R. s. rate& sale in.\\
		\hline
		Cate-1&14,279,113&20,977,374&76,508,132&0.022&102.1\\
	    Cate-2&8,219,052&9,335,687&20,034,800&0.023&209.8\\
        Cate-3&14,992,453&23,821,075&114,120,422&0.020&419.8\\
		Cate-4&14,400,948&24,659,246&131,780,136&0.026&280.7\\
		\hline
		
	\end{tabular}}
	\label{tab:data}
\end{table}
In this section, we perform experiments to comprehensively evaluate our proposed model \model{}. We aim to answer the following questions:
\begin{itemize}[leftmargin=*] 
	\item \textbf{Q1}: How does the overall performance of our model{} compared with other state-of-the-art methods in rising star prediction task? 
	\item \textbf{Q2}: Does user interest in Taocode diffusion help us with predicting rising stars?
	\item \textbf{Q3}: Does each component of our model \model{} (i.e., multi-task framework, PickGate, coupled mechanism, hierarchical dynamic structure) contribute to our model on the target task?
	\item \textbf{Q4}: Does the dynamic information contribute to our task?
	\item \textbf{Q5}: What is the effect of changing key hyperparameters such as the GNN layer size for \model{}?
\end{itemize}
\subsection{Experimental setup}
\vpara{Experiment Datasets.}
We construct four experiment datasets on the basis of different high-level categories to evaluate the effectiveness of our model. 
Due to data privacy issues, we obfuscated the exact name of each category in this paper.  
In the process of constructing each dataset, for the diffusion graph, we sample 10 thousand items that have more than one hundred Taocode sharing records from a time period of 12 weeks in 2020. 
Then, according to the sampled items, we create our four experiment datasets based on the Taocode diffusion data, Taobao purchase data and Taobao item basic information data (see details in Section~\ref{sec:Preliminaries}) during the same time period of 12 weeks in 2020.

\begin{table*}
	\caption{Rising star prediction results on four datasets}
	\resizebox{184mm}{17mm}{
	\begin{tabular}{c|ccc|ccc|ccc|ccc}
		\hline
		\multirow{2}{*}{Method} & \multicolumn{3}{c|}{Cate-1}& \multicolumn{3}{c|}{Cate-2}& \multicolumn{3}{c|}{Cate-3}& \multicolumn{3}{c}{Cate-4}  \\ \cline{2-13}
								&  Precision  &  Recall &  F1-score&  Precision  &  Recall &  F1-score&  Precision  &  Recall &  F1-score&  Precision  &  Recall &  F1-score\\  \hline
		\textbf{RiseNet}
		& 0.4072    &  0.6476 &  \textbf{0.5000}
		&0.2058 &  \textbf{0.4488} &  \textbf{0.2822}
		&\textbf{0.1911}   &  0.3093 &\textbf{0.2362} 
		& \textbf{0.1640}   &  \textbf{0.4860} &\textbf{0.2453}     
		 
		 \\ 
		DA-RNN
		& 0.3478    &  0.6038 & 0.4414
		&0.1582 &  0.2333 &  0.1886
		& 0.1429   &  0.3438 &  0.2018
		& 0.1171  &  0.3271 &  0.1724     
		 
		\\  
		LSTNet  
		& 0.2872    & 0.4297 & 0.3443
		&0.0765 &  0.3494 &  0.1225 
		& 0.0560   &  0.2110 &  0.0880
		& 0.1000   &  0.4188 &  0.1614 
		 
		\\  
		MTGNN  
		& \textbf{0.4080}    & 0.4881 & 0.4444
		& 0.0762    & 0.2892 &  0.1206 
		&  0.0743  & 0.3846 &  0.1245 
		& 0.0879&  0.4103 &  0.1448  
		 
		  \\ 
		GCN  
		& 0.2447   &  0.6571  & 0.3566
		& \textbf{0.2190} &  0.2362 &  0.2273
		& 0.1041  & \textbf{0.3918} &  0.1645
		& 0.1134   &  0.3402  & 0.1701   
		  \\   
		GIN  
		& 0.2578   &  0.6286  & 0.3657
		& 0.1638 &  0.2992 &  0.2117
		& 0.1220  & 0.3196 &  0.1766  
		& 0.0989   &  0.4845& 0.1643  
		   \\  
		DCRN
		& 0.3857 &  0.5143 &  0.4408
		& 0.1498 &  0.3858 &  0.2159
		&  0.1176  & 0.3093 &  0.1705  
		& 0.1200  & 0.3093 &  0.1729  
		  \\  
		EvolveGCN  
		& 0.2537   &  \textbf{0.6571}  & 0.3661
		& 0.1695   &  0.3937  & 0.2369 
		&  0.1352  & 0.3402 &  0.1935
		& 0.0784   &  0.3814  & 0.1301 
		  
		  \\  \hline
	\end{tabular}}
	\label{tab:overall}
\end{table*}

The overall statistics of the sampled datasets are presented in Table \ref{tab:data}, with obfuscated name of each category. 
\hide{Note that as we also introduced in the Section~\ref{sec:Preliminaries}, each rising star is labeled by its weekly updated rank in all items of its high-level category. }
We extract the temporal features of the sampled items in each time step, each of which mainly includes 
price, sales and category information. 
Moreover, the extracted features of users, who are engaged in the diffusion process, consists of 
information about  users’ purchase and diffusion habits.
The purchase information includes user consumption level, category distribution of purchase product and consumption frequency, which helps the model to predict the probability that users will buy the items in the future. The diffusion habits include users' willing to share or receive Taocodes, forwarding probability and relationship between user purchase and sharing behavior, which help the model to forecast the future diffusion network and item sales.   

\vpara{Comparison methods.}
We compare our proposed model with the following state-of-the-art methods according to various research disciplines. 

\textbf{Sales Prediction Methods:} The sales prediction task is usually considered to be a time series prediction problem. It has developed over time from utilizing conventional regression models \cite{box1970distribution}\cite{zhang2003time} to employing deep neural network~\cite{borovykh2017conditional}\cite{lai2018modeling}. Here, we considered the following state-of-the-art methods that performs well in the sales prediction domain. Note that, due to these models' inability to handle the graph data, we concatenate statistics of graph (e.g., scale, breadth) with item features as input of time series model at each time step:    
\begin{itemize}[leftmargin=*] 
	\item \textbf{DA-RNN\cite{qin2017dual}}: This is a non-linear autoregressor (AR) with attention mechanism in both encoder and decoder RNNs. Compared with A-RNN, DA-RNN assumes that the input features must be correlated over time, and uses an encoder attention to capture the dependencies.
	\item \textbf{LSTNet\cite{lai2018modeling}}: LSTNet is a deep learning framework (long- and short-term time series network), which is designed for multivariate time series prediction. This method combines a convolutional neural network and a recurrent-skip network to capture both short-term and long-term trending patterns of the time series.
	\item \textbf{MTGNN\cite{wu2020connecting}}: MTGNN is the first to address the multivariate time series forecasting problem via a graph-based deep learning approach. This method exploits the inherent dependency relationships among multiple time series.
\end{itemize}

\textbf{Graph Neural Network Models}: 
To further evaluate our model, especially the ability to quantify user interest in diffusion graph, we consider to compare \model{} with following two kinds of Graph Neural Network: 
1) Static GNN model. For these methods, we first concatenate the graph representation learnt from GNN at every time step, and then concatenate it with the item features. We pass final representation through an MLP to conduct the rising star prediction.  
2) Dynamic model. the implementation details are introduced below.
The considered state-of-the-art methods are listed in the following:
\begin{itemize}[leftmargin=*] 
	\item \textbf{GCN\cite{kipf2016semi}}: From a spatial-based perspective, GCN can be considered as averagely aggregating the feature information from a node’s neighborhood.
	\item \textbf{GIN\cite{xu2018powerful}}: GIN adjusts the weight of the central node by means of a learnable parameter in the aggregation process in order to distinguish the graph structure information.
	\item \textbf{DCRN\cite{seo2018structured}}: The DCRN model combines convolutional neural networks (CNN) on graphs to identify spatial structures and recurrent neural networks (RNN) to find dynamic patterns. We concatenate the graph representation that generated by CNN with item temporal data, and then take the concatenated vector as the input of RNN at each time step. 
	\item \textbf{EvolveGCN\cite{pareja2020evolvegcn}}:  It adapts the graph convolutional network (GCN) model along the temporal dimension without resorting to node embeddings. 
	The EvolveGCN captures the dynamism of the graph sequence through using an RNN to evolve the GCN parameters. 
	We concatenate final learnt representation of the dynamic graph and item temporal features. 
	Then the concatenated representation is passed through a MLP to predict rising star.
\end{itemize}    
 
\vpara{Evaluation Protocols.}We utilize three evaluation metrics to evaluate the performance of our model: Precision (how many positives were predicted correctly), Recall (the ability of the model to find all the positive samples), F1-score (calculates a mean of precision and recall). Considering the unbalanced nature of our labels, we pay more attention to F1-score.

\vpara{Implementation details.}Our experiment datasets contain datas that span 12 weeks in total. We use the first 8 weeks for training, weeks 5 to 10 for validating and weeks 7 to 12 for testing, so as to prevent label leakage. The utilized time steps' length for our experiment is set as four. 
The graph neural network layer number is set as two. 
We use the Adam optimizer with learning rate 0.001 to train the models. 
All the experiments are conducted on a 8-core CPU with 32 GB memory and Tesla V100 GPUs machine. 
Moreover, the algorithms are implemented in Pytorch~\cite{paszke2019pytorch} and Pytorch Geometric ~\cite{fey2019fast}.
\begin{table*}
    \caption{Ablation study of \model{} on four datasets}
	\resizebox{184mm}{15mm}{

	\begin{tabular}{c|ccc|ccc|ccc|ccc}
		\hline
		\multirow{2}{*}{Method} & \multicolumn{3}{c|}{Cate-1}& \multicolumn{3}{c|}{Cate-2}& \multicolumn{3}{c|}{Cate-3}& \multicolumn{3}{c}{Cate-4}  \\ \cline{2-13}
		&  Precision  &  Recall &  F1-score&  Precision  &  Recall &  F1-score&  Precision  &  Recall &  F1-score&  Precision  &  Recall &  F1-score \\  \hline
		RiseNet
		& 0.4072    &  0.6476 &  \textbf{0.5000}
		&0.2058 &  \textbf{0.4488} &  \textbf{0.2822}
		&\textbf{0.1911}   &  0.3093 &\textbf{0.2362} 
		&0.1640   &  \textbf{0.4860} &\textbf{0.2453}            
		 
		\\ 
		RiseNet-r
		& 0.2851   &  \textbf{0.6571} & 0.3977
		&0.1503 &  0.2047 &  0.1733
		&0.0870   &  0.3750 &0.1412 
		& 0.1111   &  0.4783 &  0.1803   
		 
		\\ 
		RiseNet-np                 
		& 0.3964    &  0.6381 &  0.4891
		&\textbf{0.2239} & 0.3543 &  0.2744
		&0.1556   &  \textbf{0.4124} &0.2260 
	    &\textbf{0.1771}   & 0.3505 &0.2353	
		
		\\ 
		RiseNet-nm                     
		& 0.3892    &  0.6190 &  0.4779
		&0.1963 & 0.3307 & 0.2463
		&0.1339   &  0.3505 &0.1937 
		& 0.0717   &  0.3608 &0.1197 
		
		\\ 
		RiseNet-nc
		& \textbf{0.4939}   &  0.3905 & 0.4362
		& 0.1838   &  0.2677& 0.2179
	    & 0.1387   &  0.3402 & 0.1970
		& 0.1523   &  0.2523 & 0.1915                    
		 
		\\
		RiseNet-nd
		& 0.2712   &  0.6095 & 0.3754
		& 0.1932   &  0.2677 & 0.2244
	    & 0.1143   &  0.3196 & 0.1685
		& 0.0803   &  0.4330 & 0.1355                  
		 
		\\\hline

	\end{tabular}
}
	\label{tab:ablate}
\end{table*}

\subsection{Overall Performance}
In Table 2, we present the overall performance of comparison methods and our proposed model \model{} on four real-world datasets in terms of Recall, Precision and F1-score. Based on these experimental results, we can answer the first research question:

1) Our model significantly outperforms other sales prediction domain methods in F1-score (improves 68.7\% in F1-score on average). This result highlights the strength of our proposed \model{} approach, which jointly
captures dynamic user interest and item features. The State-of-the-art sales prediction methods such as LSTNet and MTGNN cannot handle our target task, which can be explained by the following two reasons. First, the development process from low-turnover items to rising stars lacks periodicity, which is the key information that comparison methods such as LSTNet want to capture. Secondly, as we mentioned in Section~\ref{sec:Introduction}, items become rising stars occurs during a short time period that exhibits no long-term dependence. Therefore, MTGNN is not suitable in this case and do not perform well due to its over-fitting on long-term dependency. 

2) Compared with the state-of-the-art GNN models, our model also achieves better performance in F1-score (improves 37.3\% in F1-score on average). This demonstrates the superiority of our designed graph neural network model. For static models like GCN and GIN, their propagation methods lack the ability to model the heterogeneous influence of different Taocodes to users, which is a key to model the user interest towards items in diffusion. Also, they can not distinguish the roles of different users to retain key information when pooling node representations to graph representation. Besides, they do not capture the dynamic pattern of the evolving Taocode diffusion graphs. For the dynamic GNN models (such as DCRN and EvolveGCN), when they capture the dynamic patterns, they do not consider the dynamic interactions between item temporal features and user interest diffusion. Thus, they can't unify the two kinds of dynamic information as our model does. Besides, their GNN propagation methods are not suitable for our rising star scenario, which is the same as the above mentioned two static models. 

\subsection{Component-Wise Evaluation of \model{}}

To validate the contribution of each component of our proposed \model{}, we conduct ablation studies on our key components. The ablation experiments are listed below:
\begin{itemize}[leftmargin=*] 
    \item \textbf{Effect of User Interest in Diffusion.} 
	RiseNet-r: A simplified version of \model{} that does not include the graph neural network model (predict rising stars without the Taocode diffusion information).
	\item \textbf{Effect of PickGate.} 
	RiseNet-np: The Model is trained without the PickGate. We get the graph representation by directly apply summation function to all user embeddings.
	\item \textbf{Effect of multitask framework.} 
	RiseNet-nm: The Model is trained without the supervision of diffusion graph scale prediction task. (Graph Neural Network parameters are only updated by the loss of rising star prediction.)
	\item \textbf{Effect of the coupled mechanism.} RiseNet-nc: A simplified version of \model{} that does not include the coupled mechanism: Fuse Gate (We use a simple concatenate function to replace the Fuse Gate), and the Diffusion Gate (We remove the influence probability in the graph propagation process).
	\item \textbf{Impact of capturing temporal evolving pattern.} 
	\model{}-nd: We evaluate the effect of our dynamic model structure. We simply concatenate the learnt graph representation at every time step, and then concatenate it with item features. After getting the final representation, we use it to predict the rising stars through a MLP. 
\end{itemize}

We report the evaluation results in Table 3. It can be observed that \model{} achieves the best performance compared with all ablated model versions in F1-score. The ablation experiment results can be interpreted to suggest the following: 

1) From the comparison results with \model{}-r, our model achieves superior performance (improves 48.0\% in F1-score on average). It shows that the Taocode diffusion data that we introduced in the rising star prediction task has significantly helped us to identify rising stars with its contained user interest information.

2) The efficacy of PickGate. Without the PickGate in the graph neural network module, the performance of the model has weakened. It illustrates that by distinguishing the roles of different users, we retain more important information in the final graph representation that helps with rising star identification. 

3) The effectiveness of the multitask framework. Without the supervision of graph diffusion scale prediction task on Graph Neural Network part of model, the F1-score is obviously lower than that obtained by the full version. It shows our multi-task mechanism significantly helps our model to leverage the user interest from the dynamic diffusion graphs by adding self-supervision. 

4) The efficacy of the coupled mechanism. Removal of the coupled mechanism (FuseGate and DiffGate) yields poorer result compared with that achieved by the full version. It illustrates the effectiveness of our coupled mechanism, which captures the dynamic interplay between temporal item information and user interest information in diffusion.

5) The effectiveness of the dynamic model structure. Without the dynamic model structure, the performance of the model worsens significantly, which indicates that capturing the temporal dependency of both user interest and item features can help us to recognize rising stars. 

Based on the above ablation study results, we can now easily answer
Q2 and Q3: user interest in Taocode diffusion greatly help us with rising star prediction and each component of our model RiseNet significantly contributes to our model on the target task.


\begin{figure}[h]
	\centering{
	\includegraphics[width=0.48\textwidth]{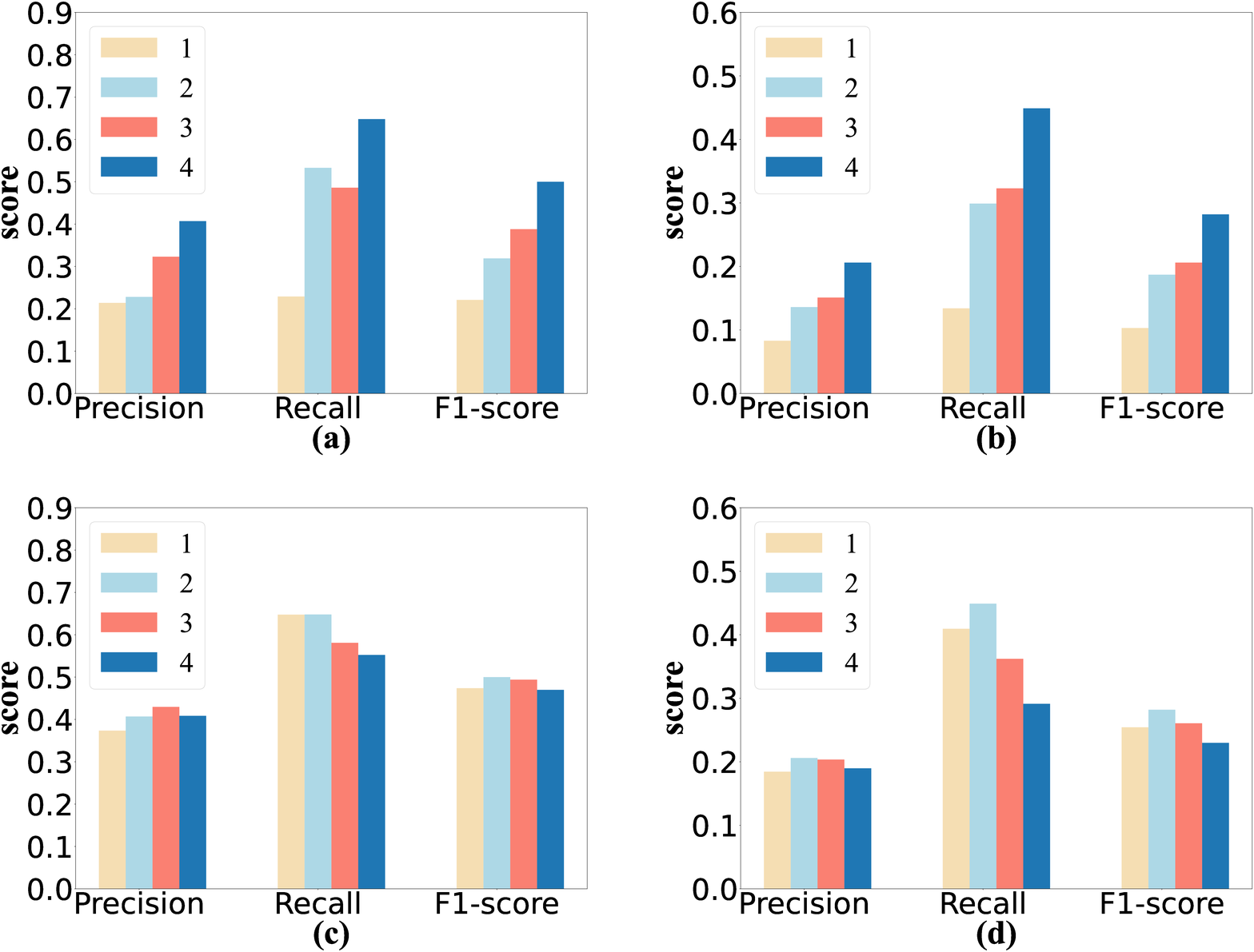}
	}
    \caption{Parameter analysis experiments of \model{}. (a) The effect of time step length on cate-1 dataset. (b) The effect of time step length on cate-2 dataset. (c) The effect of GNN layer size on cate-1 dataset. (d) The effect of GNN layer size on cate-2 dataset.}
\label{fig:exp:hyperparams}
\end{figure}
\subsection{Effect of dynamic information}
\label{sec:window}

To investigate the importance of the dynamic information, we examine the effect of utilized time steps' length on \model{}. Specifically, length one means only the last timestep's information is utilized in rising star prediction (no dynamic information). We conduct experiments on the cate-1 and cate-2 datasets with various lengths (one to six weeks). From the experimental results in Figure \ref{fig:exp:hyperparams}, we can observe that with the increase of length of time steps, the model obviously performs better in F1-score. This illustrates that utilizing the dynamic information contained in items and Taocode data could help us to recognize rising stars.

\makeatletter
\renewcommand{\@thesubfigure}{\hskip\subfiglabelskip}
\makeatother
	

	

		

\hide{
\begin{figure}[t]  
	\centering  
	
	\includegraphics[width=0.48\textwidth]{w3.eps}

	\label{fig:window}     
\end{figure} 
\begin{figure}[t]  
	\centering  
	\includegraphics[width=0.48\textwidth]{l3.eps}
	\label{fig:layer}     
	\caption{Parameter analysis of \model{} on two datasets. (a) Window size effect on cate-1 dataset. (b) Window size effect on cate-2 dataset. (c) The effect of GNN layer size on cate-1 dataset. (d) The effect of GNN layer size on cate-2 dataset.}
	
\end{figure} 
}
\subsection{Effect of GNN layer size}
\label{sec:layer}

To answer Q5, we observe the effect of GNN layer size by conducting experiments with different GNN layer sizes: one to four. From the results in Figure 5(a) and (b), we can find that increasing layer size from one to two has little improvement on the model performance in F1-score. According to our observation on the Taocode users, more than 80\% of users only have one-hop neighbor in the Taocode diffusion graph. Thus, increasing the propagation layer size cannot help us a lot to learn the user interest. Actually, applying layer size three and four even worsen the model performance compared to the result of applying layer size two, which is due to the over-smoothing problem.

\section{Related Work}
In the interests of clarity, we first review sales prediction methods, then introduce works related to dynamic Graph Neural Networks.

\vpara{Sales prediction.} 
The rising star prediction problem can be regarded as a special sales prediction problem: predicting the potential sudden increase in the future sales. The sales prediction task has played an essential role in retailing.  Traditionally, statistical methods are commonly used to predict sales; such methods include linear regression, moving average (MA), exponential smoothing \cite{hyndman2008forecasting}, Bayesian analysis \cite{yelland2010bayesian}, and so forth. 
However, statistical methods fail to capture mutual non-linear dependencies among multidimensional time series data \cite{thiesing1997sales}. Thus, deep learning models like RNNs and CNNs have shown to be effective in sales prediction. 

Borovykh etc. \cite{borovykh2017conditional} present a deep convolutional WaveNet architecture as a solution to time series predicting (sales prediction is usually regarded as a time series prediction problem~\cite{chen2018tada}). When multi-step predicting is required, the encoder-decoder architecture has been found to be helpful in real-life tasks~\cite{bao2017deep}~\cite{sordoni2015hierarchical}~\cite{bengio2015scheduled}. 
LSTNet \cite{lai2018modeling} utilizes CNN to capture short-term patterns and RNN to explore long-term dependence, which is frequently used in sales prediction. TCN \cite{bai2018empirical} combines dilated convolutions with sequence modeling. Given millions of correlated time-series, DeepGLO \cite{sen2019think} exploits both the local and global properties of time series and covariates. MTGNN \cite{wu2020connecting} proposes a graph-based deep learning approach to exploit the inherent dependencies among multiple time series.
However, most of the existing works focus on the long-term temporal prediction that has the periodicity or a regular developing trend.
In our rising star prediction task, as the sales of rising stars change drastically in a short period of time, few existing works can keep their feasibility in capturing the rising signals. To address this problem, our work propose to utilize the user diffusion information to help the rising star prediction.

Beyond the sale prediction domain, there are some works that have addressed the user interest information~\cite{hollerit2013towards}\cite{zhao2014we}. \cite{zhang2020probabilistic} proposes a probabilistic model for predicting user purchase behavior with their tweets.  \cite{zhao2015connecting} uses the user microblogging Information
in cold-Start product recommendation problem. \cite{olad2020using} utilizes tweets in product development.
However, most of these works analyse user interest information by exacting several features from their social media content and ignore the dynamic information in user social network. To the best of our knowledge, we are the first to employ the user sharing graph into the sales prediction problem.

\vpara{Dynamic Graph neural network.} 
In our framework, we use the Graph Neural Networks to model the dynamic user diffusion graph. The GNNs have attracted significant research attention over the past decades~\cite{scarselli2008graph}~\cite{kipf2016semi}~\cite{velivckovic2017graph}~\cite{ying2018hierarchical}. And as the development of static graph model, many researchers start to study the dynamic graphs \cite{zhu2016scalable}\cite{zhou2018dynamic}\cite{2018Embedding}.
Dynamic graph neural networks are designed to model the dynamic information as the graph evolving. Many real-world graphs are inherently dynamic, and their dynamic information helps to enhance the performance in many fields such as social network analysis and recommender systems. 
GCRN \cite{seo2018structured} is an extension of classical RNN on graph datasets, which exploits graphs with dynamic attributes to forecast moving MNIST data. DGNN \cite{2020Streaming} coherently captures the sequential information of edges, intervals between edges and information propagation to update node information. T-GCN \cite{zhao2019t} predicts traffic by combining GCN (where complex spatial dependence is learnt) with GRU (where temporal dependence is captured). TGAT \cite{2020Inductive} uses the self-attention mechanism and a time encoding technique to inductively infer node embeddings. The EvolveGCN \cite{pareja2020evolvegcn} captures the dynamism of the graph sequence through using an RNN to evolve the GCN parameters.
However, in our rising star scenario, the challenge lies on how to unify the dynamic information from both item and user diffusion graph information. This challenge makes the existing dynamic graph neural network unsuitable in this task. In order to solve this challenge, we propose to utilize the interplay between item and user interest with inspiration from observations. Different from the previous work, we add a special designed attention in the graph diffusion process with the utilization of dynamic information. Also, in each time step, we unify the item features and user diffusion information with the mutual dependency of them rather than simply concatenate them. This coupled mechanism enable our model to identify the rising stars more accurately.
\section{Conclusion}
In this paper, we propose to study a novel and practical problem: rising star prediction in online markets, aiming to find the promising items among the current low-turnover items. Rising star prediction benefits rational allocation of marketing resources, alleviation on unfair recommendation and balance of supply and demand. However, existing sales prediction methods are challenged by the contingency of the rising stars' sales trend. To address this, we introduce the user interest in diffusion to help with the target task and conduct a case study on Taocode diffusion data offered by Taobao. On the basis of the exploratory analysis on Taocode diffusion data, we validate the strong correlation between rising star identification and user interest diffusion. To this end, in order to leverage user interest and items features simultaneously to predict rising stars, a novel framework, named \model{}, is proposed. \model{} contains a coupled mechanism to model the dynamic interplay between user interest and items, a special designed GNN based framework to quantify the user interest and a hierarchical structure to model temporal dependency. We prove the effectiveness of our model by obtaining better performance on Taobao real-world dataset compared with other state-of-art methods.

\section*{Acknowledgments.} This work is supported by NSFC (No. 62176233) and the National Key Research and Development Project of China (No. 2018AAA0101900). 
\ifCLASSOPTIONcompsoc

\ifCLASSOPTIONcaptionsoff
\newpage
\fi

\bibliographystyle{IEEEtranN}
\bibliography{reference}

\end{document}
\endinput
